\documentclass[10pt,twocolumn,letterpaper]{article}

\usepackage{iccv}
\usepackage{times}
\usepackage{epsfig}
\usepackage{graphicx}
\usepackage{multirow}
\usepackage{amsmath}
\usepackage{xcolor}
\usepackage{amssymb}

\usepackage[breaklinks=true,bookmarks=false]{hyperref}

\iccvfinalcopy % *** Uncomment this line for the final submission

\def\iccvPaperID{50} % *** Enter the ICCV Paper ID here

\ificcvfinal\fi

\begin{document}

%%%%%%%%% TITLE
\title{Image Super-Resolution via Attention based Back Projection Networks}

\author{Zhi-Song Liu, Li-Wen Wang, Chu-Tak Li, Wan-Chi Siu and Yui-Lam Chan\\
The Hong Kong Polytechnic University\\
Hung Hom, Hong Kong\\
{\tt\small \{zhisong.liu, liwen.wang, ron.li\}@connect.polyu.hk, \{enwcsiu, enylchan\}@polyu.edu.hk}
}

\maketitle
\ificcvfinal\thispagestyle{empty}\fi

%%%%%%%%% ABSTRACT
\begin{abstract}
   Deep learning based image Super-Resolution (SR) has shown rapid development due to its ability of big data digestion. Generally, deeper and wider networks can extract richer feature maps and generate SR images with remarkable quality. However, the more complex network we have, the more time consumption is required for practical applications. It is important to have a simplified network for efficient image SR. In this paper, we propose an Attention based Back Projection Network (ABPN) for image super-resolution. Similar to some recent works, we believe that the back projection mechanism can be further developed for SR. Enhanced back projection blocks are suggested to iteratively update low- and high-resolution feature residues. Inspired by recent studies on attention models, we propose a Spatial Attention Block (SAB) to learn the cross-correlation across features at different layers. Based on the assumption that a good SR image should be close to the original LR image after down-sampling. We propose a Refined Back Projection Block (RBPB) for final reconstruction. Extensive experiments on some public and AIM2019 Image Super-Resolution Challenge ~\cite{AIM} datasets show that the proposed ABPN can provide state-of-the-art or even better performance in both quantitative and qualitative measurements.
\end{abstract}

%%%%%%%%% BODY TEXT
\section{Introduction}
\begin{figure}[t]
\vskip 0.01in
\begin{center}
\centerline{\includegraphics[width=1\columnwidth]{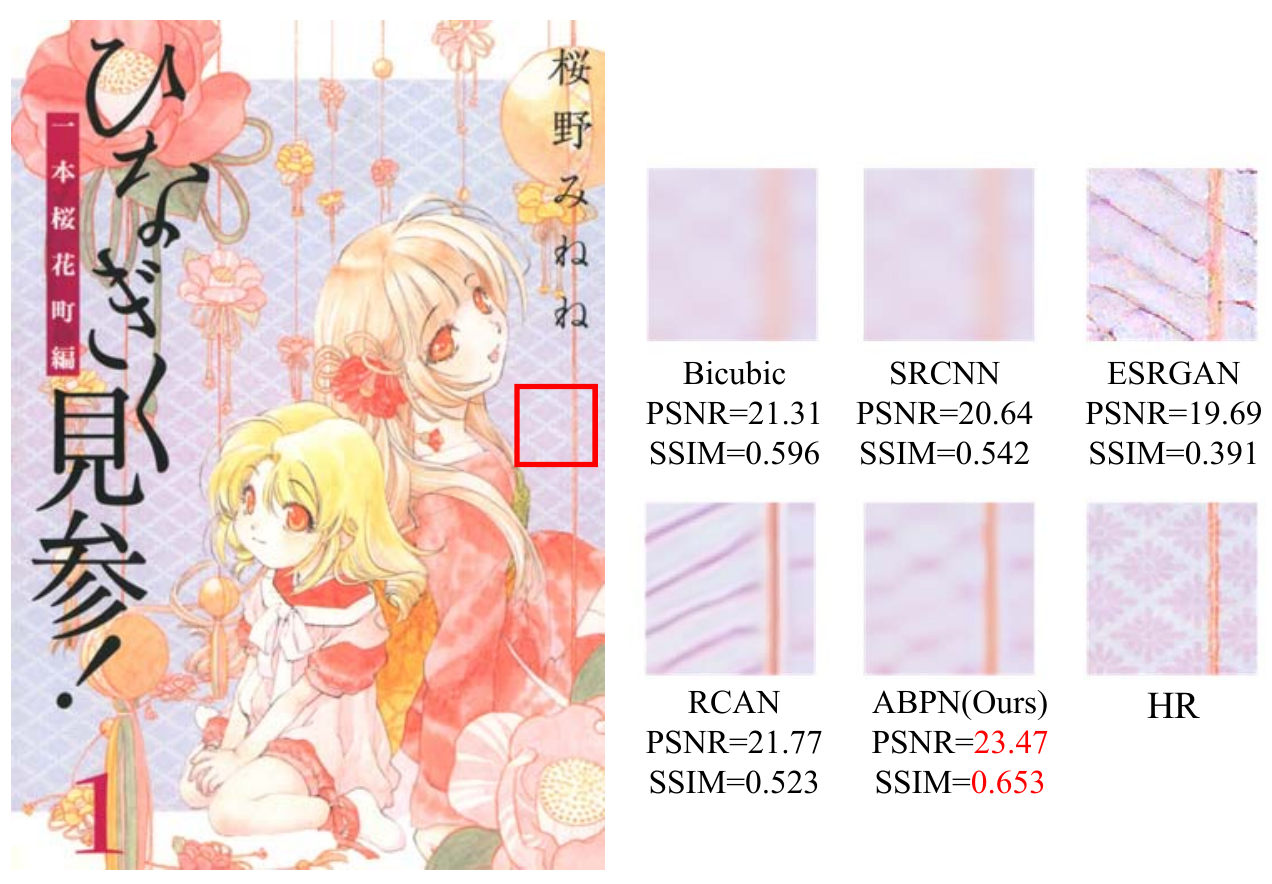}}
\caption{SR results on image \textit{HinagikuKenzan} with SR factor 16. We applied 2 times of 4$\times$ SR}
\label{Figure 1}
\end{center}
\vskip -0.3in
\end{figure}

As a fundamental low-level vision problem, image super-resolution (SR) attracts much attention in the past few years. The objective of image SR is to super-resolve low-resolution (LR) images to the desired dimension as the same high-resolution (HR) images with pleasing visual quality. For $\alpha\times$ image SR, we need to approximate $\alpha\times\alpha$ times pixels for up-sampling. Thanks to the architectural innovations and computation advances, it is possible to utilize larger datasets and more complex models for image SR. Various deep learning based approaches with different network architectures have achieved image SR with good quality. Most SR works are based on the residual mapping modified from ResNet ~\cite{ResNet}. In order to deliver good super-resolution quality, we need to build a very deep network to cover receptive fields of the image as large as possible to learn different levels of feature abstrction. The advent of 4K/8K UHD (Ultra High Definition) displays demand for more accurate image SR with less computation at different up-sampling factors. It is essential to have a deep neural network with the ability to capture long-term dependencies to efficiently learn the reconstruction mapping for SR. Attention or non-local modeling is one of the choices to globally capture the feature response across the whole image. A lot of related works ~\cite{RCAN,Sec_atten,attention,Non-local,CC,Bilateral} have been proposed for computing vision successfully. There are several advantages of using attention operations: 1) It can directly compute the correlation between patterns across the image regardless of their distances; 2) It can efficiently reduce the number of kernels and depth of the network to achieve comparable or even better performance and 3) Finally, it is also easy to be embedded into any structure for operations. As shown in Figure~\ref{Figure 1}, we tested the state-of-the-art SR approaches on 16$\times$ enlargement by applying two times of 4$\times$ SR using pre-trained models. ESRGAN~\cite{ESRGAN} and RCAN~\cite{RCAN} tend to generate fake edges which do not exist in the HR images while the proposed ABPN can still predict correct patterns.

Inspired by Non-local neural networks ~\cite{Non-local} and Back Projection based image SR ~\cite{HBPN}, we propose an Attention based Back Projection Network (ABPN) for efficient image SR. Our method focuses on studying the global feature correlation to make full use of non-local mean operation. Specifically, instead of using plain concatenation or addition operations, we propose the Spatial Attention Block (SAB) to compute the auto- and cross-correlation of the feature maps extracted at different levels. That is, we use proposed SAB to measure the similarity between two feature maps to obtain the global correlation maps. By further investigating the SR methods, we find that back projection based network is a better choice for the backbone of feature extraction because it can iteratively up- and down-sample the feature maps to update the residues of LR and HR features. To make a step forward, we propose a Refined Back Projection Block (RBPB) as the final stage to directly minimize the residues between the original LR images and down-sampled predicted SR images. 

We summarize our contributions as follows: 1) By making use of the proposed Spatial Attention Block, we modified the back projection network to Attention based Back Projection Network (ABPN) for efficient single image super-resolution. (2) We propose a Refined Back Projection Block (RBPB) to replace the common post back projection process in image SR. (3) We tested our proposed SR method on various datasets and real images. Extensive experiments show that the ABPN can achieve the state-of-the-art SR or even better performance both quantitatively and qualitatively.
%-------------------------------------------------------------------------
\begin{figure*}[h]
\vskip 0.01in
\begin{center}
\centerline{\includegraphics[width=0.8\textwidth]{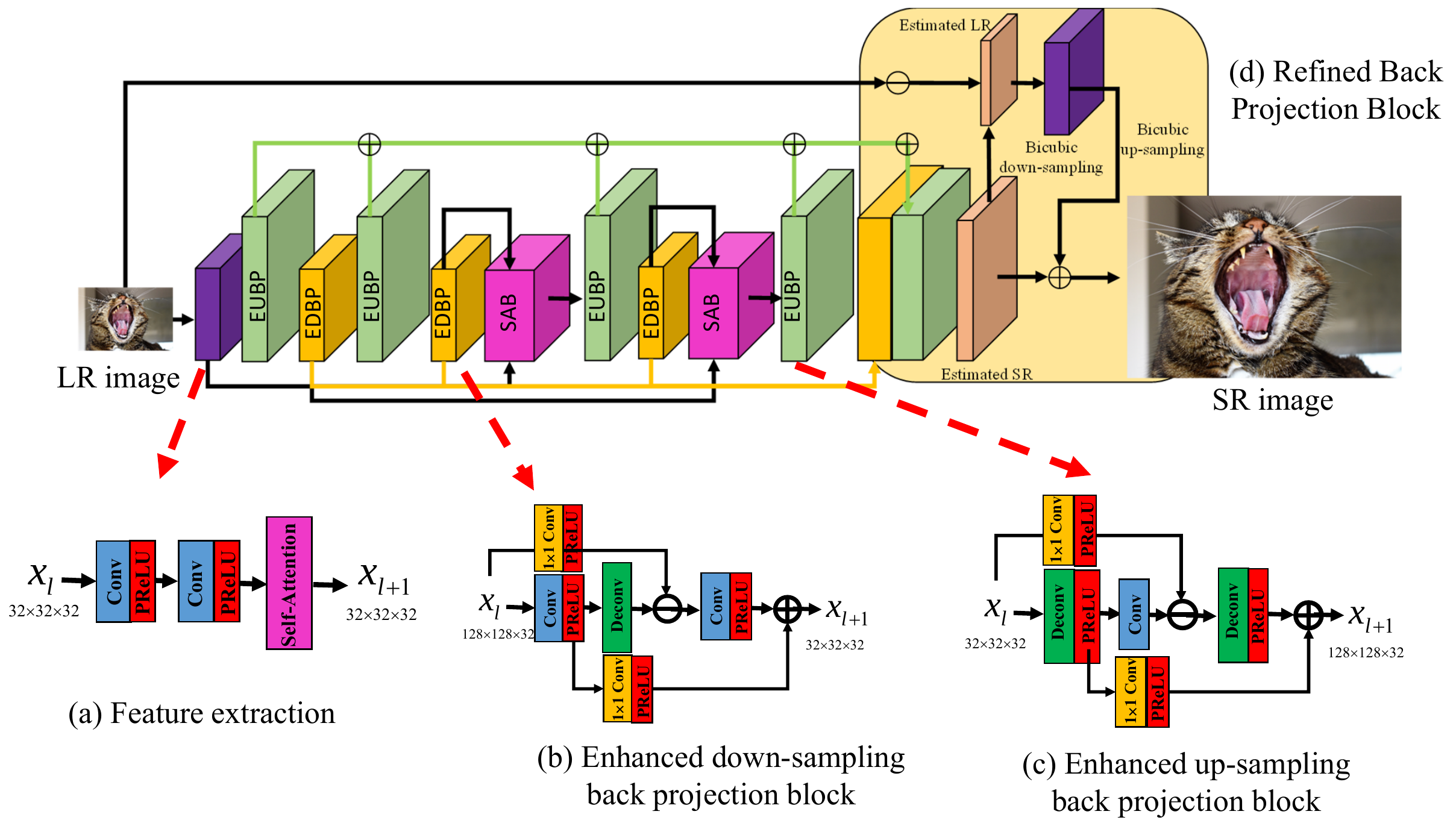}}
\caption{Proposed ABPN structure. It can iteratively up- and down-sample the feature maps to update feature residues.}
\label{Figure 2}
\end{center}
\vskip -0.3in
\end{figure*}

\section{Related Work}

\textbf{Non-local Image Processing.} Non-local mean is a conventional algorithm for image processing. The idea is that it searches not only the local areas but also the non-local areas for repeated patterns. It allows distant pixels or patches to contribute to the filtered region. The idea is generalized as a non-local convolution operation which maps the neighborhood region to the whole region of images or videos. It is commonly used in image denoising ~\cite{BM3D}, inpainting ~\cite{inpainting} and super-resolution ~\cite{Atten_SR}. 

Nowadays, non-local processing is also explicitly or implicitly embedded into deep neural networks to capture the long-term dependencies. In most deep learning algorithms, stacking more and more convolution operations with small kernels (e.g. 3$\times$3) can cover a larger receptive field for global modeling. This repeated local operation has the limitations of 1) inefficient computation for practical applications, 2) difficulty in optimizing networks and 3) a feedforward operation without feedback. Recurrent Neural Networks (RNN)~\cite{RNN} are the dominant approaches for sequential data by forming a close loop to progressively process the data. However, it  still works on a local neighborhood and its performance is not optimal. Recently, there is a trend of using self-attention~\cite{attention} or non-local neural network~\cite{Non-local} for modeling the sequential data in language and images. Note that in this paper, we use the term ``attention'' to describe the non-local modeling process in deep feature extraction. There are several great works on making use of attention mechanism in computing vision. ~\cite{attention} first proposed self-attention for machine translation. The idea is to decompose each word as a weighted combination of all positions in the sequence. That is, the model looks into onward and backward words to ensure the consistency of the translation. Similar self-attention based works were proposed in various computing fields. For example, ~\cite{Non-local} proposed non-local neural network to investigate the possible solution to spatial attention for video classification. ~\cite{CC} proposed an efficient attention computation mechanism called Criss-Cross Network for semantic segmentation. ~\cite{Bilateral} used the idea of bilateral filter to learn robust weighting model for object recognition. Besides, ``attention'' has also been proposed for image super-resolution and shown its great potential. For example, inspired by the squeeze and excitation network ~\cite{SENet}, ~\cite{RCAN} proposed to model the channel correlation by residual channel attention network. ~\cite{Sec_atten} further modified the idea of channel attention to second-order attention enhancement. However, these approaches still do not fully explore the non-local property in the spatial domain. Hence, there is a great potential for further study.

\textbf{Super-Resolution Deep Neural Networks.} In the past few years, deep neural networks have shown remarkable ability on image SR. From the beginning of the pioneer work~\cite{SRCNN}, CNN has outperformed conventional learning approaches significantly. The capabilities of resolving complex nonlinear mapping models and digestion on huge datasets encourage researchers to design deeper networks for better performance. Most of the state-of-the-art SR approaches adopt the residual architecture, like SRGAN~\cite{SRGAN}, EDSR~\cite{EDSR}, DenseSR~\cite{DenseSR} and ESRGAN~\cite{ESRGAN}. There are also some SR approaches that have different architectures for reconstruction. For example, ~\cite{pixelCNN} proposed the PixelCNN for image reconstruction. ~\cite{DRRN} proposed to use recursive neural network to iteratively predict the SR image. ~\cite{DBPN,HBPN} proposed to embed the back projection into the super-resolution to update the LR and HR feature residual. This can be considered as a generalized residual model.

Recently, using generative adversarial networks (GAN) for perceptual image SR attracts a lot of attention. The idea is to add one discriminator as an indicator for SR estimation. The backbones for generator and discriminator are more or less the same as aforementioned SR algorithms. A better architecture can further improve the perceptual quality. Once the training is finished, we only need the generator for testing. It is important to make sure the model complexity of the generator to be as small as possible for real-time applications. In this paper, we have not investigated our proposed SR method on perceptual quality but it can be modified as the generator for efficient recall.
%-------------------------------------------------------------------------

\section{Method}

\subsection{Problem Formulation}
Let us formally define the image SR. Mathematically, given a LR image $\mathbf{X}\in\mathbb{R}^{\mathit{m}\times\mathit{n}\times3}$ down-sampled from the corresponding HR image $\mathbf{Y}\in\mathbb{R}^{\mathit{\alpha m}\times\mathit{\alpha n}\times3}$, where ($\mathit{m}$, $\mathit{n}$) is the dimension the image and $\alpha$ is the up-sampling factor. They are related by th following degradation model,
\begin{small}
\begin{equation}
\mathbf{X}=\mathbf{D}\mathbf{Y}+\mu \tag{1}
\label{Equation 1}
\end{equation}
\end{small}\\
where $\mu$ is the additive white Gaussian noise and \textbf{D} is the down-sampling operator. The goal of image SR is to resolve Equation~\ref{Equation 1} as Maximum A Posterior (MAP) problem as follows,
\begin{small}
\begin{equation}
\mathbf{\hat{Y}}=\underset{\mathbf{Y}}{\arg\max}\, log \mathit{p}(\mathbf{X|Y})+log \mathit{p}(\mathbf{Y}) \tag{2}
\label{Equation 2}
\end{equation}
\end{small}\\
where $\mathbf{\hat{Y}}$ is the predicted SR image. log$\mathit{p}$($\mathbf{X}|\mathbf{Y}$) represents the log-likelihood of LR images given HR images and log$\mathit{p}$($\mathbf{Y}$) is the prior of HR images that is used for model optimization. Formally, we resolve the image SR problem as follows,
\begin{small}
\begin{equation}
\underset{\theta}{\min} \Arrowvert\mathbf{Y-\hat{Y}}\Arrowvert^\mathit{r}  \ \text{s.t.} \mathbf{\hat{Y}}=\underset{\mathbf{Y}}{\arg\min} \frac{1}{2}\Arrowvert\mathbf{X}-\mathbf{DY}\Arrowvert^2+\lambda\Omega(\mathbf{Y}) \tag{3}
\label{Equation 3}
\end{equation}
\end{small}\\
where $\Arrowvert\ast\Arrowvert^\mathit{r}$ represents the $\mathit{r}$-th order estimation of pixel based distortion. The regularization term $\Omega(\mathbf{\mathbf{Y}})$ controls the complexity of the model. Using external or internal images, we can form LR-HR image pairs to train the proposed Attention based Back Projection Network (ABPN) model to approximate the ideal mapping model. As shown in Figure~\ref{Figure 2}, the complete structure of ABPN contains three basic modules: Feature extraction, Enhanced Back Projection Blocks and Refined Back Projection Block. Feature extraction includes two convolution layers and followed by a self-attention block as a global weighting process. Enhanced Back Projection Blocks are modified from ~\cite{HBPN} and the difference are twofold: 1) the concatenation layer is replaced by the proposed Spatial Attention Block and 2), the LR feature maps are combined with HR feature map together to form the final feature maps. Finally, the Refined Back Projection Block updates the feature residues between the estimated and original LR images to refine the final SR image. The detailed structure is discussed in the following parts.

\subsection{Back Projection Blocks for image SR}
\begin{figure}[t]
\vskip 0.01in
\begin{center}
\centerline{\includegraphics[width=0.9\columnwidth]{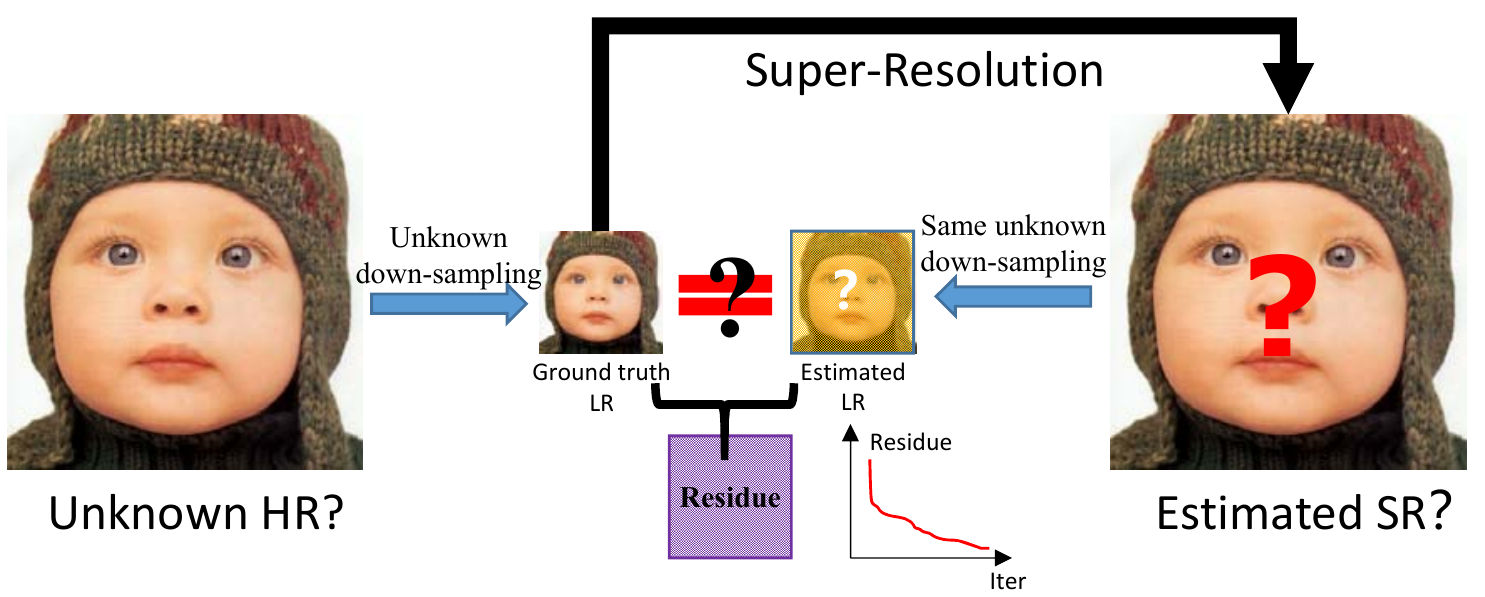}}
\caption{Back Projection procedure.}
\label{Figure 3}
\end{center}
\vskip -0.3in
\end{figure}

The Back Projection block was first proposed in DBPN ~\cite{DBPN} and the further modified version is formed in HBPN ~\cite{HBPN}. Let us see Figure~\ref{Figure 3}, the idea of back projection is based on the assumption that a good SR image should have an estimated LR image that is as close as possible to the original LR image. We follow the same idea to build our basic module entitled as Enhanced Down-sampling Back Projection blocks (EDBP) for down-sampling and Enhanced Up-sampling Back Projection block (EUBP) for up-sampling. As shown in Figure~\ref{Figure 2}, We stack multiple back projection blocks in up-down order to extract deep feature representation. For the final reconstruction, the intermediate feature maps are concatenated together to learn the SR images. The only structural difference between ~\cite{HBPN} and ours is that we also concatenate the LR feature maps together (yellow lines shown in Figure~\ref{Figure 2}) with HR feature maps for final reconstruction. Note that since the LR feature maps are $\alpha\times$ smaller than HR, we use one deconvolution layer to up-sample them to the same size as the HR feature maps.

%-------------------------------------------------------------------------
\subsection{Spatial Attention Blocks (SAB)}
\begin{figure}[b]
\vskip 0.01in
\begin{center}
\centerline{\includegraphics[width=0.9\columnwidth]{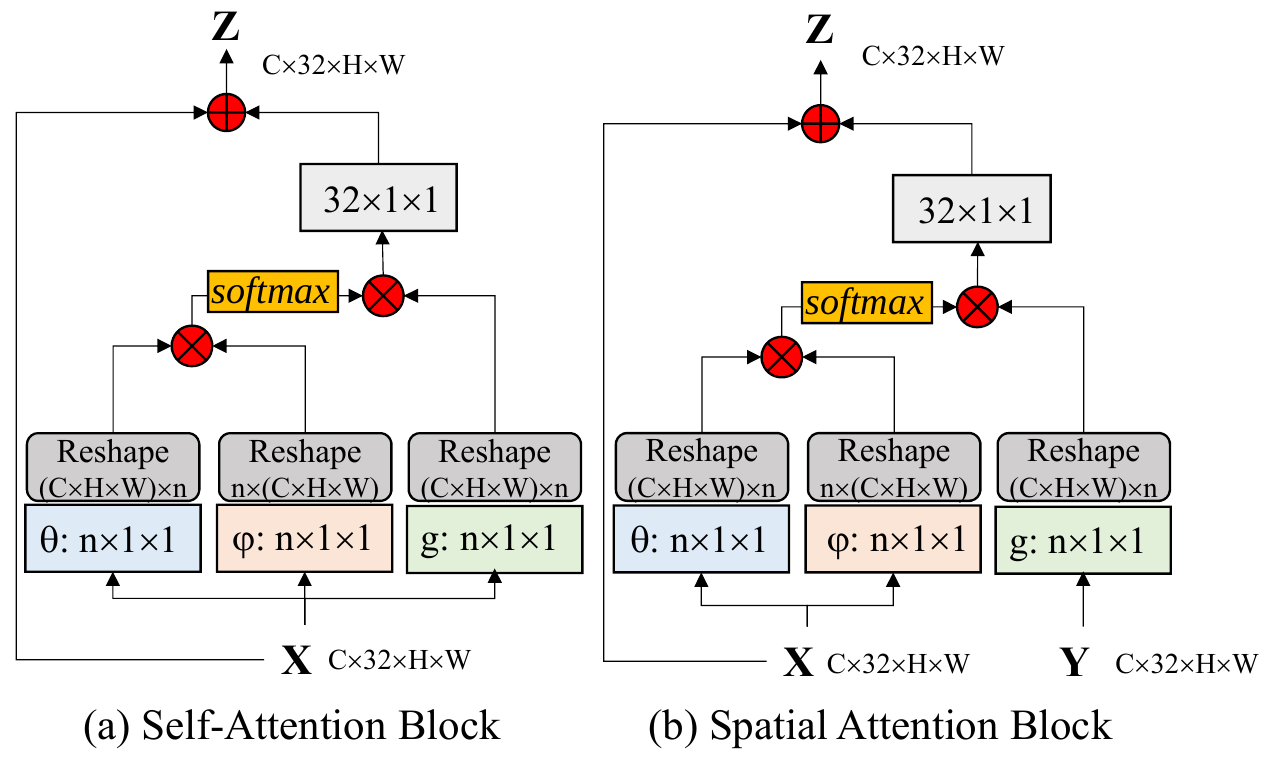}}
\caption{Comparison between self-attention and spatial attention blocks.}
\label{Figure 4}
\end{center}
\vskip -0.3in
\end{figure}

Spatial Attention Blocks are the major contribution of this work. The idea is to learn cross-correlation between features at different levels. In the proposed ABPN network, we have two types of attention blocks: self-attention blocks and spatial attention blocks. The self-attention block is exactly the same as the one in ~\cite{attention} that is situated at the end of the feature extraction (the pink block in Figure~\ref{Figure 2}(a)). And the spatial attention block is located at each EDBP block (pink blocks in Figure~\ref{Figure 2} with words ``SAB'') to extract the attention maps for following up-sampling. Their detailed differences are described in Figure~\ref{Figure 4}.

Inside self-attention and spatial attention blocks, there are three convolution layers that decompose the input data into three components: $\theta$, $\phi$ and $\mathit{g}$. Then two dot product operations are done using two of the three components. There is a short connection between input to the output so the attention models need to learn the residual mapping relationship. The difference is that the self-attention takes only the input \textbf{X} for calculation while the spatial attention block takes both \textbf{X} and \textbf{Y} for calculation. 

The attention model can be understood as a non-local convolution process. For input \textbf{X}, we can define the non-local operation as follows,
\begin{small}
\begin{equation}
\mathbf{Z}=\mathit{f}(\mathbf{X}, \mathbf{X}^T)\mathit{g}(\mathbf{X}) \tag{4}
\label{Equation 4}
\end{equation}
\end{small}\\
where \textit{f} represents the relationship of each pixel to another on the input image \textbf{X}. Following the description of self-attention, we can further rewrite Equation~\ref{Equation 4} as,
\begin{small}
\begin{equation}
\mathbf{Z}=softmax\left(\theta(\mathbf{X}\right)\phi(\mathbf{X}^T))\mathit{g}(\mathbf{X}) \tag{5}
\label{Equation 5}
\end{equation}
\end{small}\\
Similarly, for spatial attention block, we can write it as,
\begin{small}
\begin{equation}
\mathbf{Z}=softmax\left(\theta(\mathbf{X})\right)\phi(\mathbf{X}^T))\mathit{g}(\mathbf{Y}) \tag{6}
\label{Equation 6}
\end{equation}
\end{small}\\
The non-local operation in both self-attention and spatial attention consider all positions on the feature maps. The dot product of $\theta(\mathbf{X})\phi(\mathbf{X}^T)$ can be regarded as the covariance of the input data. It measures the degree of tendency between two feature maps at different channels. A convolution operation or channel attention model ~\cite{RCAN} can only sum up the weighted input in a local region while the attention model can compute the whole data, It can be also related to the Principal Component Analysis (PCA). As shown in Figure~\ref{Figure 4}, input \textbf{X} is decomposed into $\theta(\mathbf{X})$ and $\phi(\mathbf{X}^T)$. Then we vectorize the feature maps along the channel dimension so that \textit{i}-th vector represents the feature map at \textit{i}-th channel. Their dot products calculate the autocorrelation of the input data. Using Softmax operation can normalize each of the vectors to become a unit vector. Once this is done, each of the unit vector can be interpreted as an axis of the input data. Multiplying \textit{g}(\textbf{X}) to the normalized vectors can be considered as projecting data to a new coordinate system. The output of Softmax can be called the global weighting matrix that measures the importance of each feature map. Note that the goal of PCA is to reduce the dimension of data so it calculates the statistical correlation of a group of data and find the eigenvectors to project all the data with maximum variance. However, the self-attention and spatial attention focus on finding the principal features across the whole spatial domain so that they calculate the feature correlation across the channel domain and find the basis for projection. 

Generally, most deep learning based SR approaches concatenate feature maps from different layers to form a large feature map for next operation. In order to reduce the computation, a $1\times1$ convolution is used to globally weight all feature maps to output one compressed result. The disadvantage is that when the model goes deep, the more feature maps we concatenate and the heavier computation we need to cost on the $1\times1$ convolution. It is difficult to train global weighting to obtain optimal results. On the contrary, using spatial attention blocks can enhance the correlation of feature maps from different layers because the feature maps are not equally important, we only need an attention map to assign the confidence scores to the feature maps for estimation. Importantly, symbols $\theta$, $\phi$ and \textit{g} represent 1$\times$1 convolution operation without using any activation functions because 1) the correlation or covariance is a measure of linear dependence among data. Nonlinear data is more computationally demanding and 2), the input data \textbf{X} are the activated feature maps so there is no need to add another activation operation to increase the training difficulty.

\subsection{Refined Back Projection Block (RBPB)}
Finally, we have modified the Enhanced Back Projection Block to the proposed Refined Back Projection Block (RBPB) for final reconstruction. The detailed structure is shown in Figure~\ref{Figure 2}d. The reason is that the EDBP and EUBP blocks are stacked in order to update LR and HR feature residues but they never feedback to the original LR images to simulate the iterative back projection process. To form the close loop the same as Figure~\ref{Figure 3}, we use RBPB to connect the input LR image to the final SR image. In most of the SR approaches, researchers assume that the LR image is downsampled by the \textit{Bicubic} operator so we also use \textit{Bicubic} to down-sample the estimated SR image to obtain the estimated LR. Then we estimate the LR residues between estimated LR and input LR images by using another feature extraction block (the purple box at the top of Figure~\ref{Figure 2}). Finally, we up-sample the LR residues by \textit{Bicubic} and add to the estimate SR to obtain the final SR image.

\section{Experiments}
\subsection{Data Preparation and Network Implementation}
We synthesized the training image pairs based on the settings of AIM2019 SR challenge ~\cite{AIM}. The training images include 800 2K images from DIV2K ~\cite{DIV2K} and 2650 2K images from Flickr~\cite{EDSR}. Each image was rotated and flipped for augmentation to increase the number of images eight times. The LR images were obtained by using \textit{Bicubic} function in MATLAB according to down-sampling factors $\alpha$. We extracted LR-HR patch pairs from images of size 32$\alpha\times$32$\alpha$ and 32$\times$32, respectively. The testing images include Set5~\cite{Set5}, Set14~\cite{Set14}, BSD100~\cite{BSD100}, Urban100~\cite{Urban100}, Manga109~\cite{Manga109}, DIV2K~\cite{DIV2K} and DIV8K~\cite{AIM} with 4$\times$, 8$\times$ and 16$\times$ SR enlargement.

To efficiently super-resolve images, we designed the proposed ABPN network using 32 kernels for all convolution and deconvolution layers. For short connections and attention models, we used 1$\times$1 kernels with stride 1 and pad 1. For the convolution and deconvolution in EDBP and EUBP, we used 6$\times$6 kernels with stride 4 and pad 1 for 4$\times$ SR and 10$\times$10 kernels with stride 8 and pad 1 for 8$\times$ SR. Note that most SR approaches use 64 kernels for convolution or deconvolution, we only use half of convolution kernels to build the network. With the help of the proposed attention blocks, in the following experiments, we will demonstrate that the proposed ABPN can achieve comparable or even better SR performance with much less convolutional parameters.

We conducted our experiments using Pytorch 1.1, MATLAB R2016b on two NVIDIA GTX1080Ti GPUs. During the training, we set the learning rate to 0.0001 for all layer. The batch size is 8 for 1$\times10^6$ iterations. For optimization, we used Adam with the momentum to 0.9 and the weight decay of 0.0001. The executive codes and experimental results can be found in the following link:
{\color{blue}\url{https://github.com/Holmes-Alan/ABPN}}.

\subsection{Model analysis}
\textbf{Attention Back Projection Block}. For our proposed ABPN, the attention back projection block replaces the concatenation layer to combine feature maps from different layers. The self-attention is used in the feature extraction and the spatial attention is used after the enhanced down-sampling back projection blocks. To demonstrate the capability of the attention models, we design the same ABPN network using concatenation layers as \textit{Model-C} and the ABPN network using attention layers as \textit{Model-A}. Depending on the up-sampling factors, we conducted multiple experiments for 2$\times$, 4$\times$ and 8$\times$ enlargement on Set5 and Set14 to make comparison.
\begin{table}[b]
\caption{Comparison of the network using plain concatenation block or attention block, including PSNR and SSIM for scale 2$\times$, 4$\times$ and 8$\times$ SR on Set5 and Set14. {\color{red}Red} indicates the best results.}
\label{Table 1}
\vskip -0.1in
\begin{center}
\begin{small}
\scalebox{0.9}{
\begin{tabular}{c|ccccc}
\hline
\multirow{2}{*}{Algorithm} & \multirow{2}{*}{Scale} & \multicolumn{2}{c}{Set5} & \multicolumn{2}{c}{Set14} \\
 &  & PSNR & SSIM & PSNR & SSIM \\ \hline
Model-C & 2 & 37.78 & 0.955 & 33.77 & 0.913 \\
Model-A & 2 & {\color{red}38.29} & {\color{red}0.961} & {\color{red}34.18} & {\color{red}0.922} \\
Model-C & 4 & 32.48 & 0.894 & 28.78 & 0.774 \\
Model-A & 4 & {\color{red}32.69} & {\color{red}0.900} & {\color{red}28.94} & {\color{red}0.789} \\
Model-C & 8 & 26.84 & 0.774 & 24.65 & 0.618 \\
Model-A & 8 & {\color{red}27.25} & {\color{red}0.786} & {\color{red}25.08} & {\color{red}0.638} \\ \hline
\end{tabular}%
}
\end{small}
\end{center}
\vskip -0.3in
\end{table}

\begin{figure}[t]
\vskip 0.01in
\begin{center}
\centerline{\includegraphics[width=1\columnwidth]{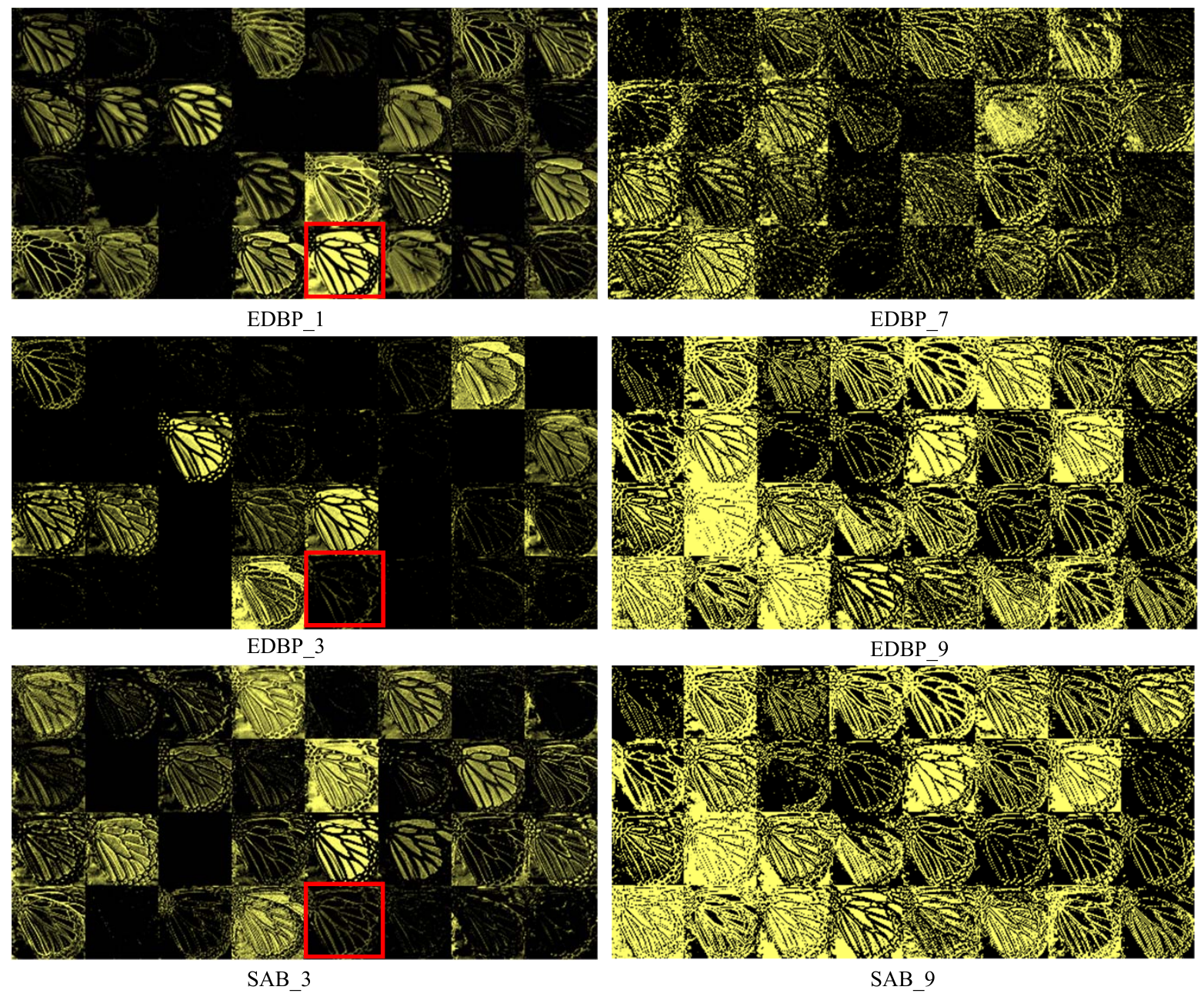}}
\caption{Visualization of the proposed spatial attention blocks. The SAB is obtained by computing the correlation between EDBP features on the first and second rows.}
\label{Figure 5}
\end{center}
\vskip -0.3in
\end{figure}

The results are shown in Table~\ref{Table 1}. We compare \textit{Model-C} and \textit{Model-A} on  SR with different up-sampling factors. \textit{Model-A} outperforms \textit{Model-C} about 0.4 dB in PSNR and 0.01 in SSIM. It indicates the effectiveness of using attention over concatenation. Furthermore, to understand the physical meaning of attention models, we visualize the feature maps obtained from EDBP and SAB blocks. The feature maps on the first row of Figure~\ref{Figure 5} were used to compute the basis for projection (same as input \textbf{X} in Figure~\ref{Figure 4}) and the feature maps on the second row of Figure~\ref{Figure 5} are projected to the basis to obtain the SAB outputs (the third row of Figure~\ref{Figure 5}). EDBP\_\textit{n} represents the \textit{n}-th down-sampling back projection blocks. NOte the red boxes on the visualization and we can find that the output of SAB blocks are the weighted results of two EDBP blocks. For example, the red boxes in EDBP\_1 are located at the feature maps that estimate the complete image so that the basis can be across the whole frequency band which shows no focus on specific features. However, the feature maps on EDBP\_3 only have responses to the edges in the neighborhood area. After the projection, the feature map on the SAB block enhanced the edge information across the whole image which is the purpose of using attention model to find the non-local property for reconstruction. 

%------------------------------------------------------------------------
\subsection{Refined Back Projection Block}

For the final reconstruction, we used the proposed Refined Back Projection Block (RBPB) to further improve the SR performance. There are some related deep learning based SR works~\cite{VDSR,CRFSR,ESRGAN} that first super-resolve the LR image via the deep networks and then use back projection as the post processing to the obtained SR image for refinement. It can improve the PSNR by about 0.01$\sim$0.1 dB but the problem is the back projection is not connected to the network to form an end-to-end architecture. We directly attached the post back projection at the end of network to jointly train the model for better SR. To make a comparison, we tested ABPN without final back projection (\textit{A}), ABPN with post back projection (\textit{B}) and ABPN with RBPB (\textit{C}) on Set5 and Set14 for 2$\times$, 4$\times$ and 8$\times$ enlargement.

\begin{table}[b]
\caption{Comparison of the network using with or without back projection or RBPB, including PSNR and SSIM for scale 2$\times$, 4$\times$ and 8$\times$ SR on Set5 and Set14. {\color{red}Red} indicates the best results.}
\label{Table 2}
\vskip -0.1in
\begin{center}
\begin{small}
\scalebox{0.8}{
\begin{tabular}{ccccccc}
\hline
\multirow{2}{*}{Algorithm} & \multirow{2}{*}{Scale} & \multirow{2}{*}{Back Projection} & \multicolumn{2}{c}{Set5} & \multicolumn{2}{c}{Set14} \\
 &  &  & PSNR & SSIM & PSNR & SSIM \\ \hline
\textit{A} & 2 & none & 38.05 & 0.960 & 33.89 & 0.919 \\
\textit{B} & 2 & post BP & 38.20 & {\color{red}0.961} & 34.07 & 0.921 \\
\textit{C} & 2 & RBPB & {\color{red}38.29} & {\color{red}0.961} & {\color{red}34.18} & {\color{red}0.922} \\
\textit{A} & 4 & none & 32.48 & 0.899 & 28.74 & 0.788 \\
\textit{B} & 4 & post BP & 32.58 & 0.899 & 28.83 & 0.788 \\
\textit{C} & 4 & RBPB & {\color{red}32.69} & {\color{red}0.900} & {\color{red}28.94} & {\color{red}0.789} \\
\textit{A} & 8 & none & 27.16 & {\color{red}0.786} & 24.97 & {\color{red}0.638} \\
\textit{B} & 8 & post BP & 27.20 & {\color{red}0.786} & 25.01 & {\color{red}0.638} \\
\textit{C} & 8 & RBPB & {\color{red}27.25} & {\color{red}0.786} & {\color{red}25.08} & {\color{red}0.638} \\ \hline
\end{tabular}%
}
\end{small}
\end{center}
\vskip -0.3in
\end{table}

The results are shown in Table~\ref{Table 2}. We can find that compared to model (\textit{A}), using back projection as a post processing for (\textit{B}) can help to boost up the PSNR performance. And when we add the Refined Back Projection Block in the network, model (\textit{C}) can further improve the PSNR about 0.1 dB. Note that the effect of back projection is limited when we super-resolve LR with larger up-sampling factors. For example, in 4$\times$ image SR, using RBPB can outperform the model without back projection by about 0.2 dB but the improvement decreases to about 0.1 dB in 8$\times$ super-resolution. The reason is that the residual information is getting smaller when the down-sampling factor is larger. Using Bicubic as the assumed down-sampling operator may not be sufficient to estimate the ground truth distribution of the LR images. 
%-------------------------------------------------------------------------
\begin{table*}[t]
\caption{Quantitative evaluation of state-of-the-art SR approaches, including PSNR and SSIM for scale 4$\times$, 8$\times$ and 16$\times$. {\color{red}Red} indicates the best and {\color{blue}blue} indicates the second best results.}
\label{Table 3}
\vskip -0.1in
\begin{center}
\begin{small}
\scalebox{0.9}{
\begin{tabular}{cccccccccccc}
\hline
\multirow{2}{*}{Algorithm} & \multirow{2}{*}{Scale} & \multicolumn{2}{c}{Set5} & \multicolumn{2}{c}{Set14} & \multicolumn{2}{c}{BSD100} & \multicolumn{2}{c}{Urban100} & \multicolumn{2}{c}{Manga109} \\
 &  & PSNR & SSIM & PSNR & SSIM & PSNR & SSIM & PSNR & SSIM & PSNR & SSIM \\ \hline
Bicubic & \multirow{9}{*}{4$\times$} & 28.42 & 0.810 & 26.10 & 0.704 & 25.96 & 0.669 & 23.64 & 0.659 & 25.15 & 0.789 \\
A+~\cite{A+} &  & 30.30 & 0.859 & 27.43 & 0.752 & 26.82 & 0.710 & 24.34 & 0.720 & 27.02 & 0.850 \\
CRFSR~\cite{CRFSR} &  & 31.10 & 0.871 & 27.87 & 0.765 & 27.05 & 0.719 & 24.89 & 0.744 & 28.12 & 0.872 \\
SRCNN~\cite{SRCNN} &  & 30.49 & 0.862 & 27.61 & 0.754 & 26.91 & 0.712 & 24.53 & 0.724 & 27.66 & 0.858 \\
LapSRN~\cite{LapSRN} &  & 31.54 & 0.885 & 28.19 & 0.772 & 27.32 & 0.728 & 25.21 & 0.756 & 29.09 & 0.890 \\
EDSR~\cite{EDSR} &  & 32.46 & 0.897 & 28.80 & 0.788 & 27.71 & 0.742 & 26.64 & 0.803 & 31.02 & 0.915 \\
RCAN~\cite{RCAN} &  & 32.63 & {\color{blue}0.900} & 28.87 & {\color{blue}0.789} & 27.77 & {\color{blue}0.744} & 26.82 & 0.809 & 31.22 & 0.917 \\
ESRGAN~\cite{ESRGAN} &  & {\color{red}32.73} & {\color{red}0.901} & {\color{red}28.99} & {\color{red}0.792} & {\color{red}27.85} & {\color{red}0.745} & {\color{blue}27.03} & {\color{red}0.815} & {\color{blue}31.66} & {\color{blue}0.920} \\
ABPN(Ours) &  & {\color{blue}32.69} & {\color{blue}0.900} & {\color{blue}28.94} & {\color{blue}0.789} & {\color{blue}27.82} & 0.743 & {\color{red}27.06} & {\color{blue}0.811} & {\color{red}31.79} & {\color{red}0.921} \\ \hline
Bicubic & \multirow{9}{*}{8$\times$} & 24.39 & 0.657 & 23.19 & 0.568 & 23.67 & 0.547 & 21.24 & 0.516 & 21.68 & 0.647 \\
A+~\cite{A+} &  & 25.52 & 0.692 & 23.98 & 0.597 & 24.20 & 0.568 & 21.37 & 0.545 & 22.39 & 0.680 \\
CRFSR~\cite{CRFSR} &  & 26.07 & 0.732 & 23.97 & 0.600 & 24.20 & 0.569 & 21.36 & 0.550 & 22.59 & 0.688 \\
SRCNN~\cite{SRCNN} &  & 25.33 & 0.689 & 23.85 & 0.593 & 24.13 & 0.565 & 21.29 & 0.543 & 22.37 & 0.682 \\
LapSRN~\cite{LapSRN} &  & 26.15 & 0.738 & 24.42 & 0.622 & 24.59 & 0.587 & 21.88 & 0.583 & 23.60 & 0.742 \\
EDSR~\cite{EDSR} &  & 26.97 & 0.775 & 24.94 & 0.640 & 24.80 & 0.596 & 22.47 & 0.620 & 24.58 & 0.778 \\
RCAN~\cite{RCAN} &  & {\color{red}27.47} & {\color{red}0.791} & {\color{red}25.40} & {\color{red}0.655} & {\color{red}25.05} & {\color{red}0.608} & {\color{red}23.22} & {\color{red}0.652} & {\color{red}25.58} & {\color{red}0.809} \\
HBPN~\cite{HBPN} &  & 27.17 & 0.785 & 24.96 & {\color{blue}0.642} & 24.93 & 0.602 & {\color{blue}23.04} & {\color{blue}0.647} & 25.24 & {\color{blue}0.802} \\
ABPN(Ours) &  & {\color{blue}27.25} & {\color{blue}0.786} & {\color{blue}25.08} & 0.638 & {\color{blue}24.99} & {\color{blue}0.604} & {\color{blue}23.04} & 0.641 & {\color{blue}25.29} & {\color{blue}0.802} \\ \hline
 &  & \multicolumn{2}{c}{DIV8K val} & \multicolumn{2}{c}{DIV2K val} & \multicolumn{2}{c}{BSD100} & \multicolumn{2}{c}{Urban100} & \multicolumn{2}{c}{Manga109} \\ \hline
Bicubic & \multirow{5}{*}{16$\times$} & - & - & 22.867 & 0.598 & 21.73 & 0.477 & 18.92 & 0.434 & 19.10 & 0.568 \\
EDSR~\cite{EDSR} &  & - & - & 24.13 & 0.631 & 22.62 & 0.506 & 19.96 & 0.481 & 20.62 & 0.635 \\
RCAN~\cite{RCAN} &  & - & - & {\color{blue}24.30} & {\color{blue}0.639} & {\color{blue}22.69} & {\color{blue}0.511} & {\color{blue}20.20} & {\color{blue}0.496} & {\color{blue}20.88} & {\color{blue}0.656} \\
ESRGAN~\cite{ESRGAN} &  & - & - & 19.09 & 0.421 & 18.01 & 0.281 & 15.42 & 0.262 & 17.41 & 0.428 \\
ABPN(Ours) &  & 26.71 & 0.65 & {\color{red}24.38} & {\color{red}0.641} & {\color{red}22.72} & {\color{red}0.512} & {\color{red}20.39} & {\color{red}0.515} & {\color{red}21.25} & {\color{red}0.673} \\ \hline
\end{tabular}%
}
\end{small}
\end{center}
\vskip -0.3in
\end{table*}

\subsection{Comparison with the state-of-the-art SR approaches}
To prove the effectiveness of the proposed ABPN network, we conducted experiments by comparing most of (if not all) the state-of-the-art SR algorithms: Bicubic, A+~\cite{A+}, CRFSR~\cite{CRFSR}, SRCNN~\cite{SRCNN}, LapSRN~\cite{LapSRN}, EDSR~\cite{EDSR}, HBPN~\cite{HBPN}, RCAN~\cite{RCAN} and ESRGAN~\cite{ESRGAN}. PSNR and SSIM were used to evaluate the proposed method and others. Generally, PSNR and SSIM were calculated by converting RGB image to YUV and only the Y-channel image was taken for consideration. During the testing, we flipped and rotated LR images for augmentation to generate several augmented inputs and then applied inverse augmentation and average all the outputs to form the final SR images. For different scaling factors \textit{s}, we excluded \textit{s} pixels at boundaries to avoid boundary effect. For these SR results, A+ and CRFSR were provided by the corresponding authors, SRCNN was reimplemented and provided by the authors of ~\cite{LapSRN}, EDSR, HBPN, RCAN and ESRGAN were reimplemented using the codes that are provided by the corresponding authors. Note that, our proposed approach also participated in the AIM2019 Image Super-resolution Challenge~\cite{AIM}. Table~\ref{Table 3} shows the comparison of all SR approaches at 4$\times$, 8$\times$ and 16$\times$. We did not conduct image SR with up-sampling factor smaller than 4 because all state-of-the-art SR approaches have achieved great performance in that scenario and the differences are too small to be compared. Instead, we show the extreme case with 16$\times$ enlargement. We chose the SR approaches that achieve the best performance in 4$\times$ and 8$\times$ for extreme comparison. The 16$\times$ results for EDSR, RCAN and ESRGAN were obtained by applying 2 times of the 4$\times$ SR using the provided pre-trained models. For a fair comparison, we also tried to use our proposed 4$\times$ ABPN SR model twice for enlargement. We can find that the proposed ABPN can achieve 0.1$\sim$0.2 dB improvement in PSNR and 0.01$\sim$0.2 in SSIM. It indicates that the proposed ABPN is more robust than others that can handle image SR even without further training. Note that we did not test Set5 and Set14 for two reasons: 1) the images in these two dataset are too small for evaluation and 2), the released codes for EDSR, RCAN and ESRGAN cannot be reimplemented in these two datasets so we tested on using DIV2K validation dataset, BSD100, Urban100 and Manga109 datasets. Furthermore, AIM2019 Image Super-resolution Challenge provided another 8K dataset for 16$\times$ SR and we show the results of using our proposed ABPN on the validation dataset. In conclusion, from the comparison on PSNR and SSIM across different up-sampling factors, we can find that using proposed ABPN can achieve comparable or even better performance compared with other state-of-the-art SR approaches. It demonstrates that the proposed ABPN is robust and accurate to handle image SR with different up-sampling factors, even in extreme conditions.
\begin{figure}[b]
\vskip 0.01in
\begin{center}
\centerline{\includegraphics[width=0.9\columnwidth]{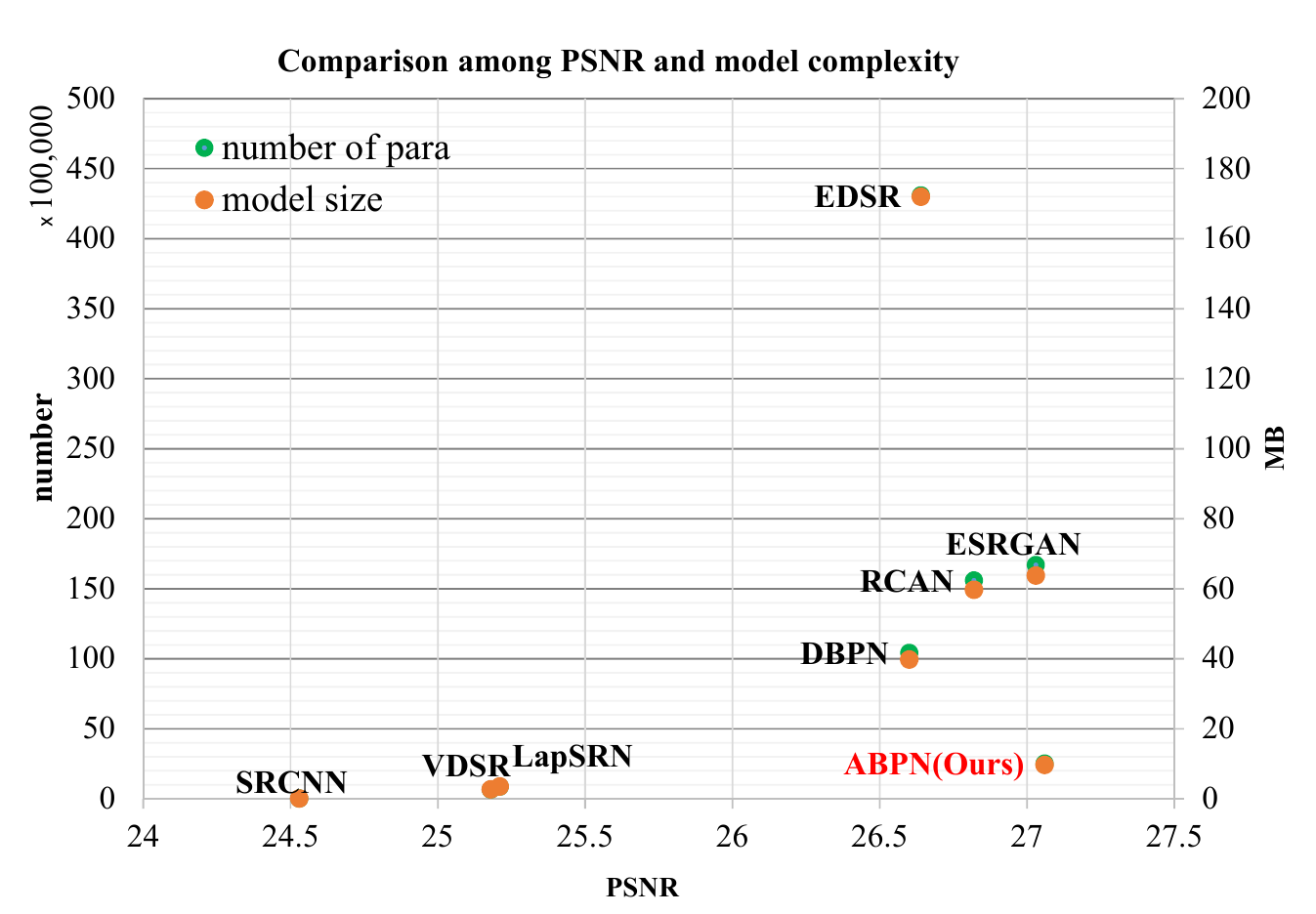}}
\caption{Comparison between model complexity and image quality. Left vertical axis is the number of parameters and right vertical axes is the size of the model file.}
\label{Figure 6}
\end{center}
\vskip -0.3in
\end{figure}

\begin{figure*}[t]
\vskip 0.01in
\begin{center}
\centerline{\includegraphics[width=1\textwidth]{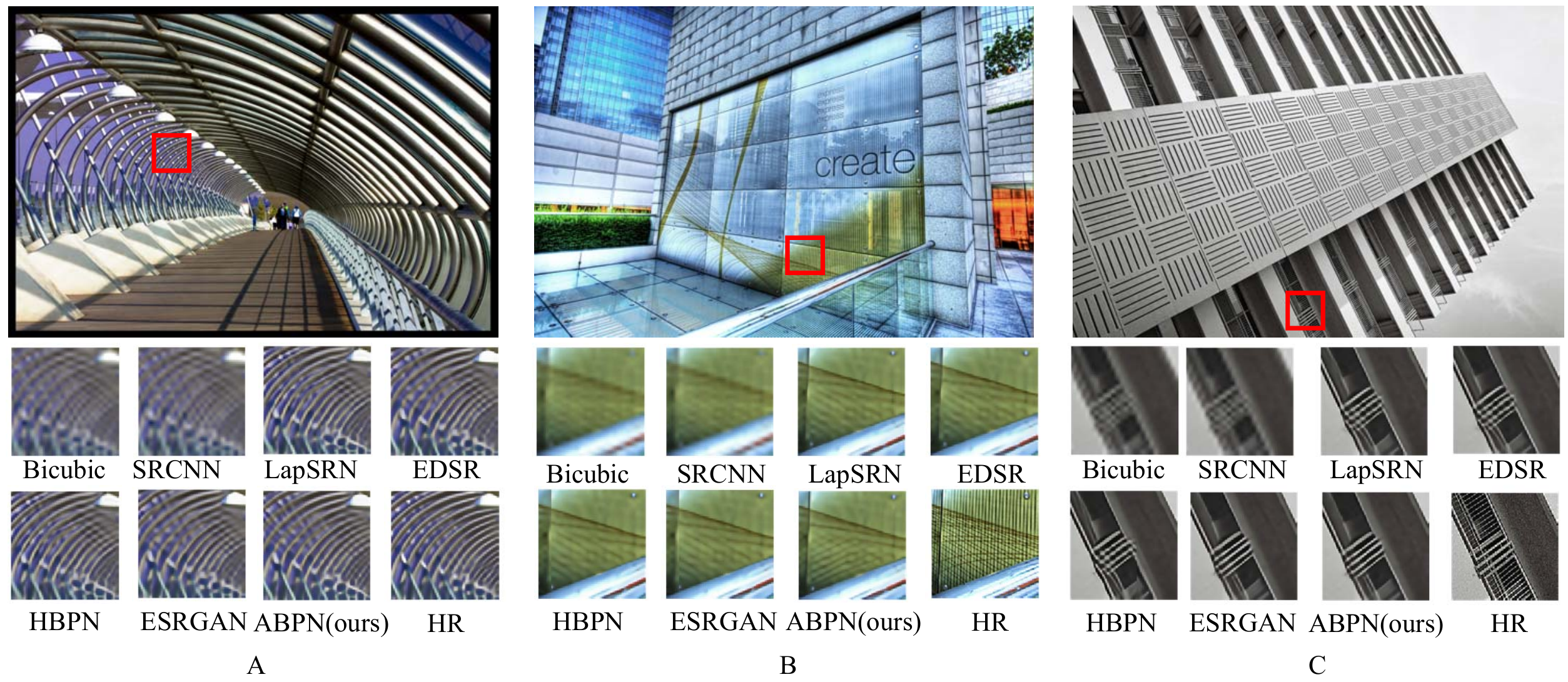}}
\caption{Visual comparison of different SR approaches on Urban100 for 4$\times$ enlargement.}
\label{Figure 7}
\end{center}
\vskip -0.3in
\end{figure*}
More importantly, we are also interested in the computation complexity of different models. Hence, we selected some of the state-of-the-art SR approaches for comparison, including SRCNN, VDSR, LapSRN, DBPN, HBPN, ESRGAN, RCAN. Note that we used the models and network setting that the authors claimed the best in their papers. We calculated the number of parameters by using the source code provided by~\cite{MSRResNet}, and used it as one indicator to show the model complexity. We also list the size of the pre-trained model file as another indicator. Since different models can be implemented with different computers and with different platforms. We did not test the running time to complicate the comparison. In Figure~\ref{Figure 6}, we show the number of parameters and PSNR for 4$\times$ SR for Urban100 dataset. 

In Figure~\ref{Figure 6}, orange dots indicate the model size and green dots indicate the number of parameters. The right bottom corner means good with higher PSNR and less model complexity. We can see that using proposed ABPN can achieve better PSNR than ESRGAN and RCAN with much less number of parameters. Note that the size of the model is consistent with the number of parameters (for some SR approaches, the orange and green dots overlap together) because the SR approaches used for comparison were all conducted using Pytorch and saved in the files with the same format. With the help of attention models, ABPN can reduce at least 2$\sim$3 times of parameters to outperform about 0.1 dB in PSNR.

Finally, we show some typical images from the testing datasets for visual comparison. Figure~\ref{Figure 7} gives the visualization of 4 $\times$ image SR. We can see that the proposed ABPN can generate SR images with comparable quality similar to other state-of-the-art SR approaches. For example, the pattern in Figure~\ref{Figure 7} B is supposed to approximately horizontal. Affected by the vertical lines on the original image, other SR approaches tend to reconstruct diagonal patterns while the proposed ABPN can correctly reconstruct the pattern. In Figure~\ref{Figure 7} C, EDSR and HBPN can generate sharp edges around the balcony but with some distortion. Our proposed ABPN can generate the pattern with better quality. 
%-------------------------------------------------------------------------
\section{Discussion}

In this paper, we explore the attention mechanism in image super-resolution, and then propose the Attention based Back Projection Network (ABPN) for image SR. There are three contributions in this network: modified enhanced back projection blocks, Spatial Attention Block (SAB) and Refined Back Projection Block (RBPB). The key modification is the Spatial Attention Block that can be used to replace the concatenation layer so that the correlation relationship between the intermediate feature maps can be extracted as a non-local weighting model. Without increasing the complexity of the CNN network, SAB can substantially improve the quality of super-resolution. The final Refined Back Projection Block works as a residual feedback that can form a close loop between the input LR and output SR images to further boost up the performance. Results on quantitative and qualitative evaluation show its advantages over other approaches. The exciting results of attention models for image SR indicate its great potential for further study. 
%-------------------------------------------------------------------------
\section{Acknowledgment}
This work was supported by the Centre for Signal Processing, Department of Electronic and Information Engineering. Earning Account, The Hong Kong Polytechnic university Internal Research Grant (ZZHR), and a RGC project of the Hong Kong Special Administrative Region, China (Grant No. PolyU 152208/17E). 
%-------------------------------------------------------------------------

{\small
\bibliographystyle{ieee_fullname}
\bibliography{paper_review}
}

\end{document}